# From Citation count to Argumentation count: a new metric to indicate the usefulness of an article.

Hardik Joshi
Asst. Professor
Department of Computer Science, Gujarat University, Ahmedabad, India
hardikjjoshi@gmail.com

**Abstract**
Citation count is a quantifiable measure to indicate the number of times an article is cited by other articles. It is believed that if an article is cited often then it must be an important or influential article; however, there is no guarantee that the most cited articles are good in quality. In this paper, the author suggests "argumentation count", a new metric for citation analysis. The proposed metric, argumentation count is a triplet of quantities for each concept of an article that helps in providing a quantifiable measure about the usefulness of an article.

**Introduction**
Since many years, most of the publication house, academic search engines and academia are using citation count as a quantifiable measure to indicate the impact or usefulness of an article. A citation count is the number of times an article is cited by other articles. It is usually considered to indicate the quality of the article. If an article is cited often it must be an important or influential article. It is widely used as a benchmarking tool to rank the articles or journals (David 2002).

Citation analysis or bibliometrics is a field of study that deals with the importance and usage of articles, journals and author profiles. Citation count is a quantifiable metric that can be applied to an article, journal or an author. Citation count can be applied to:
- an individual article to signify how often it was cited
- an author to signify the total citations, or average citation count per article
- a journal to signify average citation count for the articles in the journal

In recent days, new metrics and tools are gaining importance that is based on citation count. The h-index tries to categorize output of a researcher (Hirsh 2005). Improved metrics are being proposed over the earlier ones for instance g-index was proposed over h-index (Leo 2006), however, most of the metrics are based on the idea of citation count.

Most of the academic search engines are providing citation count along with the search results. It is assumed that the citation count is always less than the actual count as the search engines may not be able to consider each article that has cited the work. A more serious issue is that the citation count cannot identify the authenticity of the article in terms of quality and usefulness. It is always assumed that an article is of importance if it is highly cited. Few researchers try to take advantage of citation analysis by using self-citation to manipulate the citation count.

With the development of information retrieval technologies (Manning and Raghvan 2010), a new term called "argumentation mining" is being conceptualized that can be reasonably used to identify the importance of an article. The subsequent section of this paper explores the term argumentation mining and advocates its use as a metric for bibliometrics.

## Argumentation Mining

Argumentation deals with the process of construction and handling arguments. Argumentation is a part of human intelligence and is applied in day-to-day use. There is a wide span where argumentation can be found. Argumentation is mostly found in legal documents, product reviews, parliamentary speeches etc. Argumentation is generally used to analyze the pros and cons of any situation or an object. Literally works also deal with argumentation where an author tries to emphasize his ideas on the basis of the earlier ideas with proper argumentation.

Argumentation mining is a relatively new term that deals with argumentation. Argumentation mining aims to detect the arguments presented in a text document, the relations between them and the internal structure of each individual argument (Palau, Mochales, and Moens 2009). Argumentation mining deals with the detection of all the arguments involved in the argumentation process, their individual or local structure, i.e. rhetorical or argumentative relationships between their propositions, and the interactions between them. Researchers have investigated methods for argumentation mining of legal documents (Mochales and Moens 2011; Bach et al. 2013; Ashley and Walker 2013; Wyner et al. 2010), on-line debates (Cabrio and Villata 2012), product reviews (Villalba and Saint-Dizier 2012; Wyner et al. 2012), and newspaper articles and court cases (Feng and Hirst 2011).

## Argumentation Count

The idea of argumentation count is borrowed from two different fields: argumentation mining and citation count that is used in bibliometrics. By and large, argumentation count is an extension to citation count. The proposed metric "argumentation count" can help in identifying the usefulness and trust-worthiness of an article and the concepts of an article. The proposed metric can be divided into two different categories:
- Feature/Concept identification
- Argumentation triplet for each feature/concept

Each article can be assumed as a set of features or concepts. Within an article, elements like premises, hypothesis, conclusions, data, claims etc. are required to be identified for each article. Any article 'A' can be represented as a set of argument elements, 'E'. An article can be represented using the following notation:
$A = \{E_1, E_2, ..., E_n\}$

With each element, an argumentation triplet (F,N,D) is associated. Each citation of an element within an article accounts to vote for single attribute of the triplet. A triplet consists of any of the three votes, viz. Favoring or Neutral or Disfavoring. The triplet can take form as below:
{Favoring, Neutral, Disfavoring}

In general, the argumentation count of any article can be viewed as a sum total of the triplets of each of the elements. The final argumentation count can be represented as:
$AC = \{E_{1(F,N,D)}, E_{2(F,N,D)}, ..., E_{n(F,N,D)}\}$ .

For instance, argumentation count for a research article that makes two claims identified by claim-1 and claim-2 respectively can be $\{Claim\text{-}1_{(23,3,4)}, Claim\text{-}2_{(2,1,15)}\}$. By looking at the argumentation count, it can be concluded that claim-1 is cited to be favorable by 23 authors while claim-2 is cited to be disfavoring by 15 authors.

## Challenges in Argumentation Count

On being a relatively new area, there are hardly any tools and technologies that support argumentation mining. Few of the possible challenges are listed below:

- Automatic identification of argument elements from the article(e.g., premises and conclusion; data, claim and warrant), argumentation schemes, relationships between arguments in a document, and relationships to discourse goals (e.g. stages of a "critical discussion") and/or rhetorical strategies
- Creation/evaluation of argument annotation schemes, relationship of argument annotation to linguistic and discourse structure annotation schemes, (semi)automatic argument annotation methods and tools, and creation/annotation of high-quality shared argumentation corpora
- Processing strategies integrating Natural Language Processing methods and Artificial Intelligence models developed for argumentation mining

## Bringing Argumentation Count in practice

Lot of research has begun in the area of argumentation mining. Experts from various fields like computer science, library science, linguistics etc. are working in the area of argumentation mining. Although the research progress in this area is in its infancy, new tools and techniques are being developed to take up the challenge. To develop the tools and technologies for argumentation count, a good corpus of article database is required where each element within an article is identified and tagged to avoid disambiguation of automatic element identification.

As the concept of argumentation count is an extension to citation count, few issues like self-citation and coverage of adequate number of research articles while generating the citation count may creep into the calculation of argumentation count. However, the proposed metric of argumentation count can help to give a reasonable good metric for the usefulness of any article on the basis of identification of argument elements.